\documentclass[10pt]{wlscirep}
\usepackage[utf8]{inputenc}
\usepackage[T1]{fontenc}

\usepackage{graphicx}
\usepackage{dcolumn}
\usepackage{bm}

\usepackage[colorinlistoftodos]{todonotes} 

\usepackage[normalem]{ulem}
\usepackage[utf8]{inputenc}
\usepackage{amssymb}   
\usepackage{amsmath}   
\usepackage{braket}    
\usepackage{subfig}    
\usepackage{comment}   
\usepackage{float}	   
\usepackage{multirow}
\usepackage{color}
\usepackage{fancyhdr}
\usepackage{amsthm}
\usepackage{tabularx}
\usepackage{setspace}
\usepackage[export]{adjustbox}
\usepackage{stfloats}
\usepackage{tabu}
\usepackage{xcolor}

\newcommand{\be}[1]{\begin{equation}\label{#1}}
\newcommand{\ee}{\end{equation}}
\newcommand{\bea}[1]{\begin{eqnarray}\label{#1}}
\newcommand{\eea}{\end{eqnarray}}

\newcommand{\Fig}[1]{Fig.(\ref{#1})}

\newcommand{\Eq}[1]{Eq.(\ref{#1})}

\usepackage{fancyhdr}
\title{Hong-Ou-Mandel multiplexer and switch using \\ parallel chains of non-identical micro-ring resonators}

\author[1,*]{Peter L. Kaulfuss}
\author[2]{Paul M. Alsing}
\author[1]{Richard J. Birrittella}
\author[2]{A. Matthew Smith}
\author[2]{James Schneeloch}
\author[3]{Edwin E. Hach III}
\affil[1]{Booz Allen Hamilton, 8283 Greensboro Drive, McLean, VA 22102}
\affil[2]{Air Force Research Laboratory, Information Directorate, 525 Brooks Rd, Rome, NY 13411}
\affil[3]{School of Physics and Astronomy, Rochester Institute of Technology, 85 Lomb Memorial Drive, Rochester, NY 14623}

\affil[*]{kaulfuss\_peter@bah.com}

\begin{abstract}
Micro-Ring Resonators (MRRs) allow us to access the Hong-Ou-Mandel (HOM) effect at a variety of tunable parameter combinations along exact analytic solutions. This higher-dimensional space of parameters for which the HOM effect occurs constitutes what is known as a Hong-Ou-Mandel manifold (HOMM). Using a parallel series of non-identical MRRs and changing relative round-trip phase shifts between MRRs allows for the manipulation of the wavelength locations of the HOM effect. Through clever design and fabrication, we can mold the HOMM to place multiple HOM effects, or lack thereof, precisely at desired locations in wavelength. In this paper we discuss how to adjust non-identical MRR parameters to change the resulting HOMM. We also promote example designs that exhibit advantageous HOMM structures, and highlight some of the diverse possibilities that can be accessed with different circuit design. Our main examples are: 1) a wavelength division multiplexer example that matches the HOM effect locations with the already established channels to integrate with a classical communication network and 2) a HOM-based entanglement switch that allows for the rapid switching between 2-photon NOON state outputs and completely separable single photon outputs.
\end{abstract}
\begin{document}

\flushbottom
\maketitle
%
%
\thispagestyle{empty}

\section*{Introduction}

The Hong-Ou-Mandel (HOM) effect \cite{HOM:1987} is a fundamentally quantum interference phenomenon prevalent in a wide range of quantum applications. The HOM effect is of vital importance in quantum computing \cite{Kok:2007,KLM:2001,Alsing_Hach:2019,Okamoto:2011} \cite{Bouchard_2021}, quantum communication, and cryptography \cite{Gisin:2002,Pirandola:2019,Hofmann:2012,Narla:2016} because it is an irreducible two-photon interference phenomenon in which initially separable pairs of identical photons may become entangled following passage through respective input ports of a 50/50 beamsplitter. The HOM interference visibility is used to measure the indistinguishability of photons from a variety of sources such as: quantum dots \cite{Sanaka:2009,Flagg:2010,Patel:2010,Wei:2014,Senellart:2017}, atomic vapors \cite{Felinto:2006,Chaneliere:2007,Yuan:2007,Yuan:2008,Chen:2008}, nitrogen-vacancy centers in diamond \cite{Bernien:2012,Sipahigil:2012,Sipahigil:2014}, molecules \cite{Kiraz:2005,Lettow:2010}, trapped neutral atoms \cite{Beugnon:2006,Specht:2011}, and trapped ions \cite{Maunz:2007} \cite{Kaulfuss:2023Identical,Bouchard_2021}. Specifically, in quantum information processing \cite{EPR:1935,Ekert:1991,Dowling:2008,Giovannetti:2011}, two-photon entanglement is an extremely important resource and often the HOM effect is a simple and useful tool to generate and manipulate two-photon entanglement and assess photon indistinguishability.

In most applications, the HOM effect is currently achieved using beam splitters. While this is a robust and simple way to achieve the HOM effect, we build upon previous works that offer the MRR as an attractive alternative to the beam splitter for achieving the HOM effect \cite{Hach:2014,Kaulfuss:2023Identical}. In integrated photonics, MRRs are commonly fabricated and used in various ways, including as on-chip tunable beam splitters, as a means of resonantly enhancing biphoton sources, and in sensing applications \cite{Kues:2017}. Since the MRR as a component is already frequently used in integrated photonics, its use in demonstrating an enhancement to achieving the HOM effect is another advancement that will be important as we move towards circuits entirely on-chip and packaged from source to detection.

The increased tunable parameters of the MRR allow for an entire manifold of parameter solutions where the HOM destructive interference condition can be met, i.e. the 
probability for the
destructive interference of the quantum amplitudes for  the $\ket{1,1}$ state at the output is zero,
$P_{1,1}=0$ \cite{Hach:2014, HOM:1987, Alsing_Hach:2019, Kaulfuss:2023Identical}. By comparison, there is only one value of the reflectivity, $r=0.5$, for which a beam splitter will produce the HOM effect. In this paper we extend this argument and show the benefits gained by designing, fabricating, and using linear chains of non-identical parallel MRRs to achieve the HOM destructive interference condition in much more complex and useful structures than a beam splitter could ever provide. The strength of these non-identical MRR-based devices is the ability to precisely place the wavelengths where the HOM condition is satisfied through a combination of device fabrication and tunability. The MRRs are tunable via heaters placed on the ring. A voltage can be applied to the heater to change the effective index of refraction of the MRR and therefore the round-trip phase shift as a result \cite{Chen:07,Serafini:20}.

The main advantage of a series of non-identical MRRs is the increased control and manipulation of the features of the HOMM. By adjusting the number of MRRs in the series and the round-trip phase shift between MRRs in the series, one has almost complete control over the placement of the Hong-Ou-Mandel effect locations. Effectively there is extreme control over the programming of the structure of the linear transformation that occurs due to these devices.

Each MRR acts as a `tine' in a `comb' structure of spikes (in the HOMM parameter space)
that have HOM dips (in frequency) between them. Our two main applications to showcase some of the possibilities are a 5 MRR multiplexer with evenly-spaced `crescent' HOM structures that would allow for entanglement distribution across a network and a HOM-based entanglement switch using spike structures to allow for rapid changes between completely separable outputs and a two-photon NOON state output. 

At this point, we would be remiss not to mention that in addition to MRRs, Mach-Zehnder Interferometer (MZI) modules are also ubiquitous in integrated photonics, and may also define manifolds of parameter solutions where the HOM destructive interference condition can be met. However, concatenating series of MZIs cannot reproduce the same comb structure at the same level of device complexity of the corresponding MRR devices in this work because of the lack of recursive loops in the paths that light can take through an MZI-based system, which otherwise define the narrow resonant spikes of the MRRs. For this reason, we focus our attention only on MRR-based HOMM devices as a means to customize the specific frequency bands at which the system will exhibit HOM interference.

The overview of the paper is as follows. First we give a brief overview of the HOM effect in MRRs. In the Results section we describe the effect of non-identical MRRs in linear chains and show how the parameters can be adjusted to fabricate a device that meets the requirements of the system for control of wavelength placement of the HOM effect. We show a `HOM Multiplexer' and our main HOM multiplexer example with a 5 MRR device. We describe a very interesting application of linear chains of non-identical MRRs, an `HOM-based Entanglement Switch', which allows for the rapid switching between a completely separable state output to a maximally path entangled state output. Finally, we summarize our results and draw conclusions on the utility of linear chains of non-identical MRRs. In the Supplementary Material, we show in section A the effect of $2\pi$ increments to swap between spike and crescent structures in the HOMM; and show in section B an example of loss in a non-identical MRR device and discuss the asymmetry of the loss.

\section*{Background}
\label{Background}

The transformation matrix of a beam splitter that imparts a $\pi$-phase between reflected modes is given by 
\begin{equation}
    \text{BS} = 
    \begin{pmatrix}
        \cos{(\phi/2)} & \sin{(\phi/2)} \\
        -\sin{(\phi/2)} & \cos{(\phi/2)}
    \end{pmatrix},
\end{equation}
where $T=\cos^{2}\tfrac{\phi}{2}$ and $R=\sin^{2}\frac{\phi}{2}$ are the beam splitter transmittance and reflectance, respectively, and where $\phi=\pi/2$ represents a 50:50 beam splitter. 

The HOM effect occurs when a pair of indistinguishable photons are incident on the two inputs of a 50:50 beam splitter at the same time \cite{HOM:1987}. The general input/output relation for a beam splitter with a $\ket{1,1}$ input is
\begin{equation}
    \ket{1,1}_{\text{in}}\xrightarrow{}\cos{\left(\phi\right)}\ket{1,1}_{\text{out}} 
    + \sin{\left(\phi\right)}\ket{\psi_{2002}}_{\text{out}},
\end{equation}
where $\ket{\psi_{2002}}_{\text{out}}=\frac{1}{\sqrt{2}}\left(\ket{2,0}_{\text{out}}-\ket{0,2}_{\text{out}}\right)$ is the 2-photon NOON state.  When the beam splitter is 50:50, $\cos{\phi}=\cos{\tfrac{\pi}{2}}=0$ and the $\ket{1,1}_{\text{out}}$ term drops out, resulting in the 2-photon NOON state. 

The MRR can, in its simplest form, be thought of as an extended beam splitter, as shown in \Fig{fig2diagram}:

\begin{figure}[H]
	\centering
    \includegraphics[width=0.25\linewidth,keepaspectratio]{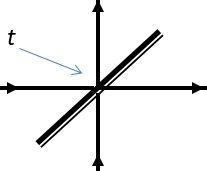}
	\includegraphics[width=0.25\linewidth,keepaspectratio]{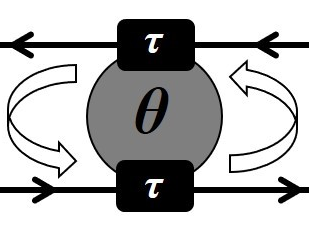}
	\caption{A diagram highlighting the change from a beam splitter to a double-bus Micro-Ring Resonator (db-MRR). The beam splitter can be described using a single transmission coefficient, t. The db-MRR with identical couplings can be described using two parameters: $\tau$ to describe the coupling coefficients and $\theta$ to describe the round-trip phase shift.}
	\label{fig2diagram}
\end{figure}

Where the MRR has two directional couplers, each is described by a coupling coefficient $\tau$, and a round-trip internal phase shift, $\theta$. Using these tunable parameters, $\tau$ and $\theta$, there are many combinations of $(\tau,\theta)$ that yield the same effect as a 50:50 beam splitter \cite{Hach:2014}. Therefore, even a single MRR is able to access the HOM effect for a larger solution set than the single point (50:50) for a beam splitter. We call this higher dimensional manifold of solution points the Hong-Ou-Mandel Manifold (HOMM) \cite{Kaulfuss_thesis:2021,Kaulfuss:2023Identical}.

\begin{figure}[H]
    \centering
    \begin{tikzpicture}[scale=1]
        \input{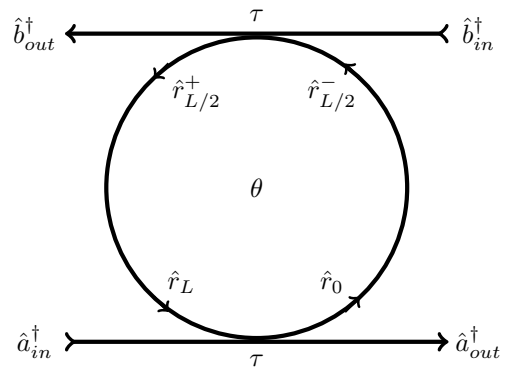}
    \end{tikzpicture}
    \caption{Diagram of all the modes in a single double-bus MRR.} 
    \label{Single MRR diagram}
\end{figure}

To model a single MRR we define modes at critical points along the path of the photons, then use a boundary condition technique to solve for the output modes as functions of the input modes. The interior modes are necessary to fulfill the system of equations, but ultimately drop out of the final results. We label the  inputs and outputs as: $\hat{a}_{in}$, $\hat{a}_{out}$, $\hat{b}_{in}$, and $\hat{b}_{out}$. We also label interior modes of the MRR, designated by $\hat{r}$, and create four interior mode locations: $\hat{r}_0$, $\hat{r}_{L/2}^{-}$, $\hat{r}_{L/2}^{+}$, and $\hat{r}_L$. 
The interior modes are labeled by subscripts indicating their location (distance $z\in[0,L]$)
inside the MRR (measured in the counter clockwise direction) as shown 
in  \Fig{Single MRR diagram}.  

\noindent The interior modes are required so that at each $\tau$ junction we have a BS-type interaction with two modes entering and two modes exiting the BS. The modes $\hat{r}_0$ and $\hat{r}_L$ are the interior modes immediately inside and just before exiting the MRR for the lower waveguide. Similarly, $\hat{r}_{L/2}^{-}$ and $\hat{r}_{L/2}^{+}$ are the modes just inside and just before exiting the MRR for the upper waveguide. 
Phase accumulation within the MRR occurs, in general, as 
$\hat{r}_{z} = e^{i \theta (z-z_0)/L}\hat{r}_{z_0}$ for $z-z_0>0$. Once the boundary condition system of equations is solved, the interior modes will be eliminated in order to obtain the algebraic relationship between the input and output modes directly \cite{Kaulfuss:2023Backscatter}.

We simply use $\tau$ to represent the coupling parameters between the MRR and the waveguide at each directional coupler. The value of $\tau$ represents the fractional transmission amplitude for a photon to continue through the waveguide instead of coupling into the MRR. Given $\tau$, we can find $\kappa$ using the reciprocity relation of the directional coupler: 
\begin{equation}
    \kappa=\sqrt{1-\tau^2}.
\end{equation}

Thus, we define the unitary directional coupler transfer matrix between input and output modes as:
\begin{equation}
\left(
\begin{array}{cc}
\tau & \kappa \\
-\kappa & \tau
\end{array}
\right). \label{BSmatrix}
\end{equation}
Applying this transfer matrix to our input/output combinations at each directional coupler yields the following equations:
\begin{alignat}{3}
    \hat{a}_{out} &=\tau \, \hat{a}_{in} + \kappa \, \hat{r}_L,  \;\;\;\;\;\;\;\;\;\;\;\;\;\; \hat{r}_0 &&=-\kappa \, \hat{a}_{in} + \tau \, \hat{r}_L, \nonumber \\
    & \\
    \hat{b}_{out} &= \tau \, \hat{b}_{in} + \kappa \, \hat{r}_{L/2}^{-},\;\;\;\;\;\;\;\;\;  \hat{r}_{L/2}^{+} &&=-\kappa \, \hat{b}_{in} + \tau \, \hat{r}_{L/2}^{-}. \nonumber
\end{alignat}

We also create equations to relate the internal modes, including the internal phase propagation of the MRR, where $\alpha$ is the phenomenological round-trip amplitude loss coefficient \cite{Yariv:2000}. Loss is introduced via the following equations: 

\begin{equation}
    \hat{r}_{L/2}^{-}=\alpha e^{i \frac{\theta}{2}}\;\hat{r}_0,\;\;\;\;\;\;\;\;\; 
    \hat{r}_L=\alpha e^{i \frac{\theta}{2}} \, \hat{r}_{L/2}^{+}. 
\end{equation}

However, for most of the results in this paper, we will assume the lossless scenario, setting $\alpha=1$.

Using a boundary condition solve method we are able to use the above coupled-mode equations to obtain solutions for $\hat{a}_{out}$ and $\hat{b}_{out}$ in terms of $\hat{a}_{in}$ and $\hat{b}_{in}$, forming a $2\times 2$ unitary transfer matrix for the single MRR, involving the MRR transfer matrix elements $A_1$, $A_2$, $B_1$, and $B_2$. In order to propagate quantum states through the MRR equation, we follow the work of Skaar, et al. (2004) \cite{Skaar:2004}, rearranging our matrix equation to obtain the transition amplitudes (in terms of the creation operators) for the input modes as a function of the output modes:

\begin{equation}
    \begin{pmatrix}
    \hat{a}_{in}^\dagger \\
    \hat{b}_{in}^\dagger
    \end{pmatrix}
    =
    \begin{pmatrix}
    A_1 & B_1 \\
    A_2 & B_2
    \end{pmatrix}
    \begin{pmatrix}
    \hat{a}_{out}^\dagger \\
    \hat{b}_{out}^\dagger
    \end{pmatrix}.
\end{equation}
Here we have assumed the constants $(\kappa,\tau,\alpha)$ are constant over the bandwidth of the photons being sent. This implicitly places us in the regime we will assume throughout the paper, where the input photons are both identical and nearly monochromatic, being distinguished only by differing input spatial modes.

The probability of a coincident output (one photon in each output mode) with a single photon input on each input mode will be written as $P_{1,1}$ and is the square of the permanent of the $2\times 2$ transfer matrix. This is the probability that we will concern ourselves with for all the results in this paper. 

\begin{equation}
    P_{1,1}=|A_1\,B_2+A_2\,B_1|^2 \label{PABAB}
\end{equation}

The HOM effect occurs when $P_{1,1}=0$. In the case of MRRs, this equation yields the manifold of solutions we called a HOMM \cite{Kaulfuss:2023Backscatter}. We assume for the cases in this paper that the single photon inputs are indistinguishable in frequency and arrive simultaneously. For a single MRR with identical directional couplers, the contour plot of the HOMM is shown in \Fig{1MRRcontour}, where the red line represents the exact 1-D HOMM: 

\begin{figure}[H]
    \centering
	\includegraphics[width=0.5\linewidth,keepaspectratio]{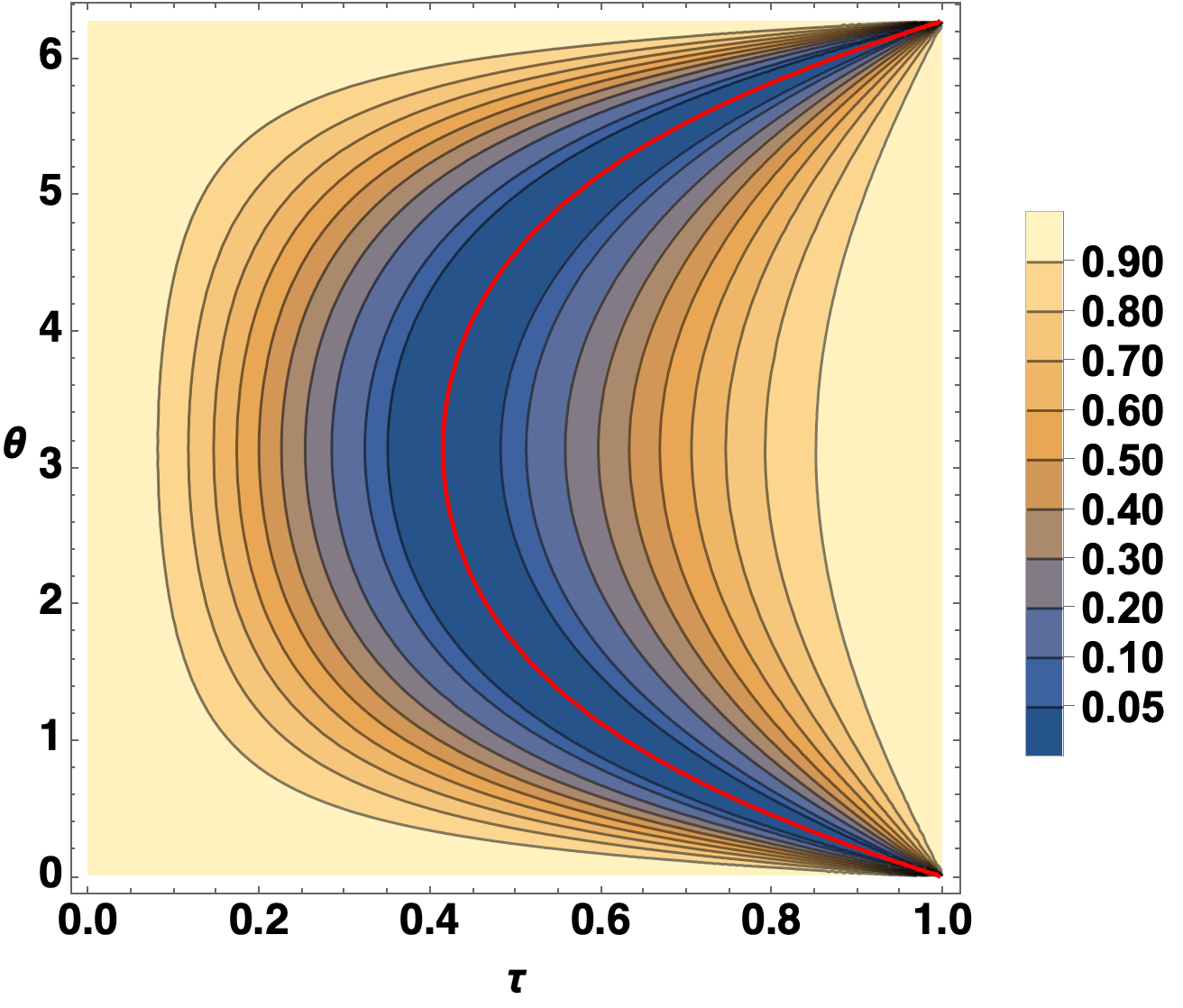}
	\caption{Single MRR Contour. Contours are of $P_{1,1}$ where the darkest blue region is $P_{1,1}<0.05$. This contour plot legend applies for all contour plots in this paper. The red line in this contour plot represents the exact HOMM. }
	\label{1MRRcontour}
\end{figure}


This enhancement to the accessibility of the HOM effect can be furthered by creating parallel chains of multiple MRRs. We first explored identical chains of MRRs, where each MRR in the chain has the same $\tau$ and $\theta$ values. This yields a set of exact analytic solutions for the HOMM depending on $N$, the number of identical MRRs in the parallel chain \cite{Kaulfuss:2023Identical}:

\begin{equation}
    \tau_N(\theta)=(2-\cos{\theta}-\sqrt{3-4\cos{\theta}+\cos^2{\theta}})^{\frac{1}{2N}}
    \label{exactidentical}
\end{equation}

Identical parallel chains of MRRs are extremely stable around the optimal HOM location ($\tau=\tau_{\text{min}},\theta=\pi$) \cite{Kaulfuss:2023Identical}. The trade-off for stability is that these identical chain HOMM solutions are only based on $N$, the number of identical MRRs in the chain. The identical chain creates a single crescent shaped HOMM curve according to \Eq{exactidentical} and does not allow for any further control over the placement of the HOMM (other than changing $N$). In this paper, we illuminate the advantage gained by allowing the MRRs in the parallel chain to be non-identical. This allows for extreme control over the shape of the manifold because each MRR creates its own feature on the overall HOMM. Using different numbers of MRRs in the chain and adjusting the round-trip phase shift increments allows the HOM effect  to be achieved at any desired locations via precise base fabrication or further dynamic tuning. This is the advantage we explore how to harness in this paper.

\section*{Results}
\label{Results}

The Hong-Ou-Mandel Manifold (HOMM) \cite{Hach:2014,Kaulfuss:2023Identical} is a higher-dimensional manifold where the HOM condition is met $P_{1,1}=0$, where $P_{1,1}$ is the probability of finding the $\ket{1,1}$ state, exemplified in \Eq{PABAB}). When using linear chains of non-identical MRRs there is broad control of the features of the HOMM.   Careful design and fabrication of MRR-based devices can create a wide variety of HOMM shapes to fit different applications. Put simply, each additional MRR added to the chain is an additional opportunity that can be used to introduce another (typically spike shaped) structure to the overall HOMM.

As shown in \cite{Hach:2014} a single MRR yields, in general, a 9-dimensional solution HOMM if all parameters are left free (4 complex coupling parameters and a round-trip phase) and each additional MRR in the chain could similarly contribute 9 additional dimensions. Therefore, circuits with larger chains of MRRs produce a generally larger space of solutions in the manifold that yields the HOM effect. In practice, the complex phases on each of the four coupling parameters are trivial, and by identically coupling both waveguides to each MRR in the circuit we are able to characterize the system using a single, real, coupling parameter, $\tau$, and the round-trip phase shift of the first MRR in the chain, $\theta$. Each MRR in the chain beyond the first has constant increments added or subtracted to the round-trip phase shift of the first MRR, which we will refer to as the ``base" MRR. Still, the total number of structures that can be manipulated in the HOMM of the non-identical chain is always equal to the total number of MRRs in that chain, because each MRR can have a different round-trip phase shift. 

For a full, in-depth, treatment of loss in MRRs see one of our previous papers \cite{Alsing_Hach:2017a,Kaulfuss:2023Backscatter}. The effect of loss on these systems is a decrease in the overall probability of detecting both photons at the output ports. The location of the HOM effects given by the manifold remain unchanged, but the visibility of the HOM dips would be decreased. Visibility is an important quantity defined in \Eq{Visibility} that we use to qualify our ``HOM Switch'' examples. Small amounts of loss do not have a large effect on the HOMM, so we will ignore the effects of loss for the remainder of these results and assume the circuits to be near lossless. We will discuss loss in more detail in the supplementary material.

For our examples in this paper we will always flatten the parameter space to a single $\tau$ and base $\theta$ shared across all MRR(s) leaving us with a 2-D HOMM.  This results in a vertical cross section of the space in figures such as Fig.(1). This also means that all the MRRs are identically coupled to the waveguides and the round-trip phase shift of each MRR is characterized by reference to the base MRR. The spacing of the structures on the HOMM is controlled by the relative round-trip phase shifts between the successive MRRs in the chain. The base MRR that all others in the chain are compared to is usually seen as a crescent shape in the HOMM (as seen in \cite{Kaulfuss:2023Identical}). Each additional MRR in the circuit with an offset to the round-trip phase shift of the MRR is seen initially as a spike structure, whose placement can be precisely controlled using the round-trip phase shift offsets. These offsets are equivalent to changing the size of the MRR in design and fabrication (See \Eq{lambdaeq} for the relationship between the phase offsets, photon wavelength, and MRR design parameters). 

\begin{figure}[H]
	\centering
	\includegraphics[width=0.5\linewidth,keepaspectratio]{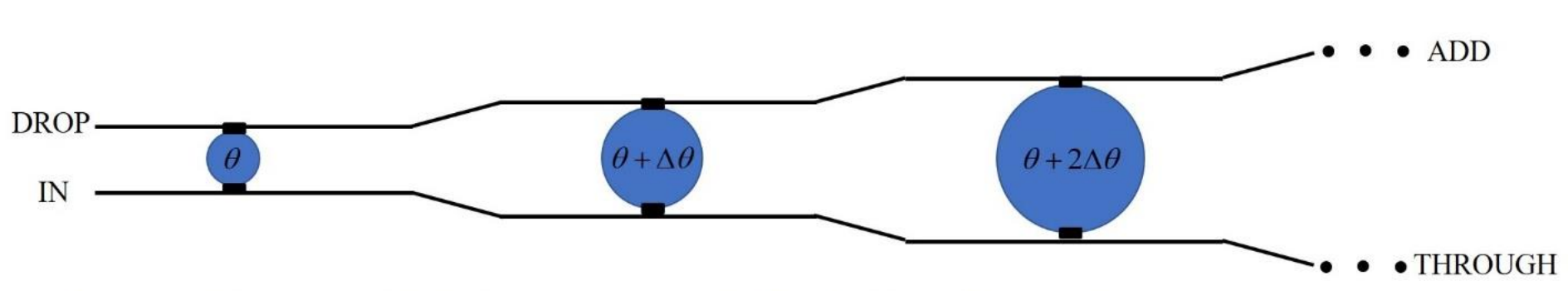}
	\caption{A linear chain of identically coupled non-identical double-bus Micro-Ring Resonators (MRRs). This device can be optimized to produce the two-photon NOON state, a signature of the HOM effect, at specific, tunable, and dynamically adjustable wavelengths.}
	\label{fig1diagram}
\end{figure}

We work
under the assumption that all of the MRRs in a given chain are identically coupled to the
waveguides accessing them; mathematically we accomplish this by assuming a single, real
transmission amplitude, $\tau$, for each of the directional couplers (small black rectangles) represented
in \Fig{fig1diagram}.

The equation relating the wavelength of the photon to the number of round trips taken in the MRR, the radius of the MRR, and the index of refraction of the material (silicon) is given as \cite{Rabus:2007}: 

\begin{equation}
    \label{lambdaeq}
    \lambda=\frac{4\pi^2n_sR}{\theta}
\end{equation}

If we take the index of refraction for silicon to be $n_s=3.48$ and assume that the radius of a base MRR with no additional increment is $25.52$$\mu$m, we find that input photons with a wavelength of $1550$nm yield a $\theta_0$ of approximately $720\pi$. Recall that for an MRR the resonance condition is $\theta=2\pi m$, where $m\in\mathcal{Z}^{+}$. This resonance-to-resonance distance is known as the free-spectral range (FSR). Adding one FSR to the round-trip phase shift, i.e. $\theta=\theta_0+2\pi=722\pi$, yields a wavelength of approximately $1545.7$nm. Thus the distance between our base resonance and the next resonance, the $0$ to $2\pi$ range in round-trip phase shift, can be thought of as a range of $1550$nm to $1545.7$nm in wavelength, meaning the FSR is approximately $5$nm due to the resonance condition of the MRR for these values. 

\begin{figure}[H]
    \centering
	\includegraphics[width=0.5\linewidth,keepaspectratio]{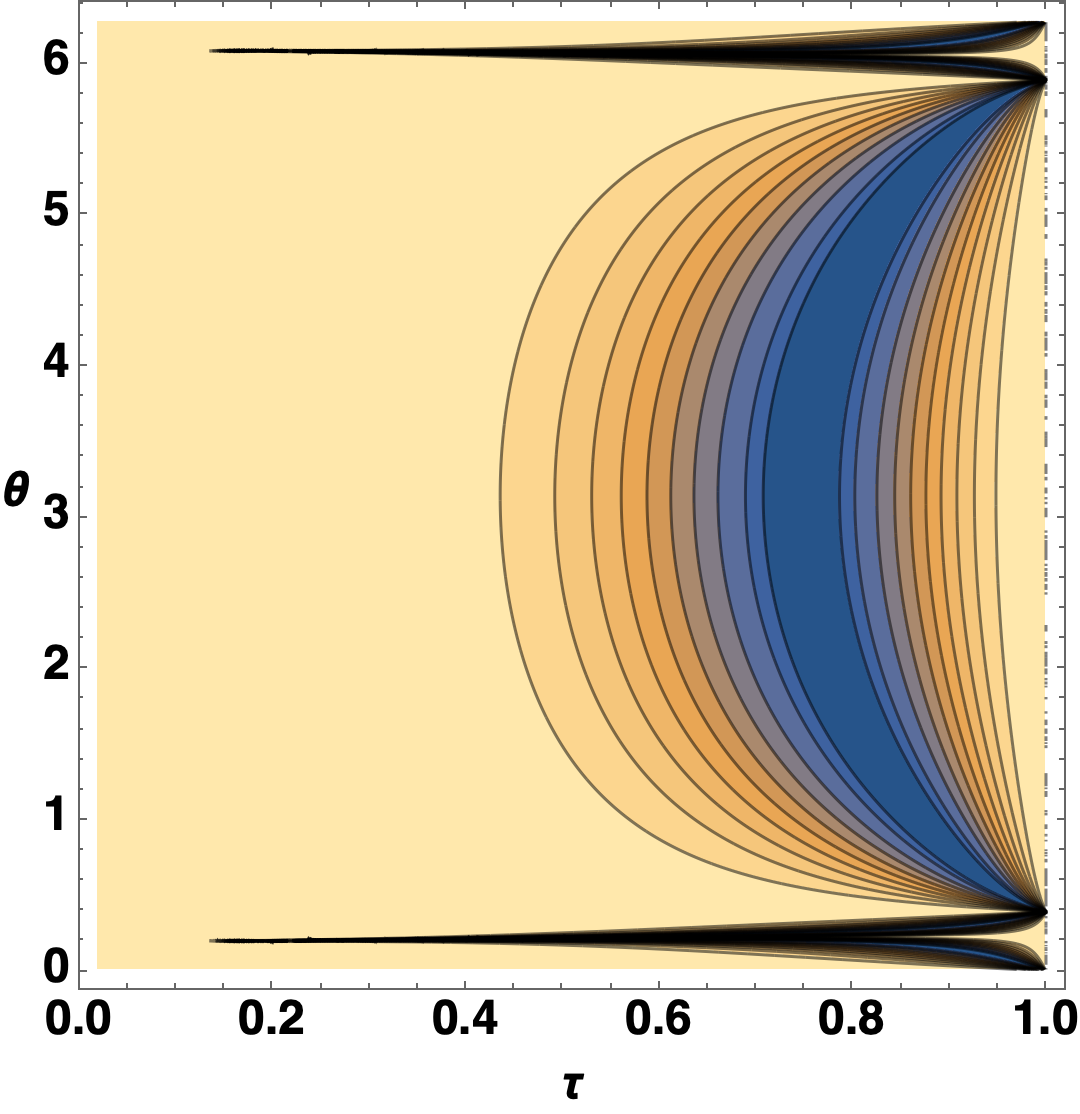}
	\caption{3 MRR Contour Plot with increments: [$-\frac{\pi}{8},0,\frac{\pi}{8}$]. We refer to the structures created on the edges by the non-identical MRRs as `spikes' and the main structure between the spikes as a `crescent'.}
	\label{3mrrexample}
\end{figure}

Using theoretical techniques \cite{Hach:2014,Alsing_Hach:2017a,Kaulfuss_thesis:2021,Kaulfuss:2023Backscatter,Kaulfuss:2023Identical}, we plot in \Fig{3mrrexample} the output photon
coincidence probability as a function of base MRR round-trip phase shift, $\theta$, and coupling parameter, $\tau$.
The two-dimensional HOMM, $\tau(\theta)$, along which $P_{1,1}=0$, exists within the darkest blue `valley' in the contour plot shown in \Fig{3mrrexample}. This darkest blue region is where the HOM effect is experimentally viable with joint-detection probabilities $P_{1,1}\leq 0.05$. In principle, we could write out an analytical result for the exact HOMM of any of these cases, just as we did for the linear chain of identically coupled identical MRRs \cite{Kaulfuss:2023Identical}. However, the result here is cumbersome and of less use. In the mode of operation that we are proposing for the category of circuit we are analyzing, the detailed knowledge of the HOMM is not required in any form. Instead, we are proposing devices that `sample’ that manifold at resolvable and controllable points where the HOM condition is met.

Negative offsets in the round-trip phase shift compared to the main crescent caused by the base MRR place spikes below the main crescent whereas positive offsets place spikes above. The structure of the MRR causes the pattern to be repeated cyclically with a $2\pi$ period. It is important to keep in mind that each of the structures caused by additional MRRs only occupies the size of the gap between successive MRRs in relative phase shift. For example, a three MRR series with relative phase shifts: $\{-\frac{\pi}{8},0$,$\frac{\pi}{8}\}$ will have a spike covering the phase range from 0 to $\frac{\pi}{8}$ at the start of the $2\pi$ cycle, a spike covering the phase range from $\frac{15\pi}{8}$ to $2\pi$ at the end of the $2\pi$ cycle, and a main crescent that covers the entire phase range between the two spikes caused by the base MRR, as seen in \Fig{3mrrexample}. Conversely, by making the MRR chain design evenly spaced in phase offset, we can cause each structure to occupy the same amount of relative phase range, shown in \Fig{5mrrcontour}. The choice of $\tau$ value to slice across also influences not only the height of the spike (in probability), but also the width in $\theta$ of the feature. Due to the flexibility and the number of tunable parameters (even for this 2-D HOMM), it is possible to create a device (based on both the number of MRRs in the chain and the relative round-trip phase shifts between them) to shape the peaks and troughs of a HOMM that meets the HOM condition at whatever locations fit the requirements of the device for a specific application.

In this paper we use two main structures when constructing our Hong-Ou-Mandel Manifolds: spikes and crescents. Spikes are the base-case when additional non-identical MRRs are added to the chains and the size of the spike in $\theta$ space is determined by the gap between the phase offsets of the corresponding MRRs. Spikes can be converted to crescents by manipulating the phase offsets of the MRRs to include additional $2n\pi$ increments where n is an integer. We showcase the utility of crescents in the following HOMWDM example and similarly we show a use case of spikes in the HOM-based Entanglement Switch example.

\subsection*{HOMWDM}
\label{WDMsec}

The first and main class of devices we explore utilizing a parallel-chain design of non-identical MRRs is what we call a ``HOM Multiplexer''. The chain of MRRs can be constructed in a way to turn all the spike structures into evenly-spaced crescents. This HOM multiplexer has a comb-like structure with evenly spaced `tines' where the HOM condition is NOT met. Spaces between the tines of the comb allow for the HOM effect to be present at specific wavelengths (frequencies), while the wavelengths between will yield separable states. The locations of the tines can be adjusted using different numbers of non-identical MRRs in the chain and different increments between the successive round trip phase shifts of the rings as seen in \Fig{fig1diagram}. 

One of the natural extensions of the multiplexer structure is as a HOM filter. A broadband signal could be sent through the device and, acting as a filter, the device would produce separable states for most wavelengths and only produce the HOM effect at a specific chosen wavelength. This effectively filters the HOM effect to a single specific wavelength allowing all other wavelengths of light to simply pass through the device. 

\begin{figure}[H]
    \centering
	\includegraphics[width=0.5\linewidth,keepaspectratio]{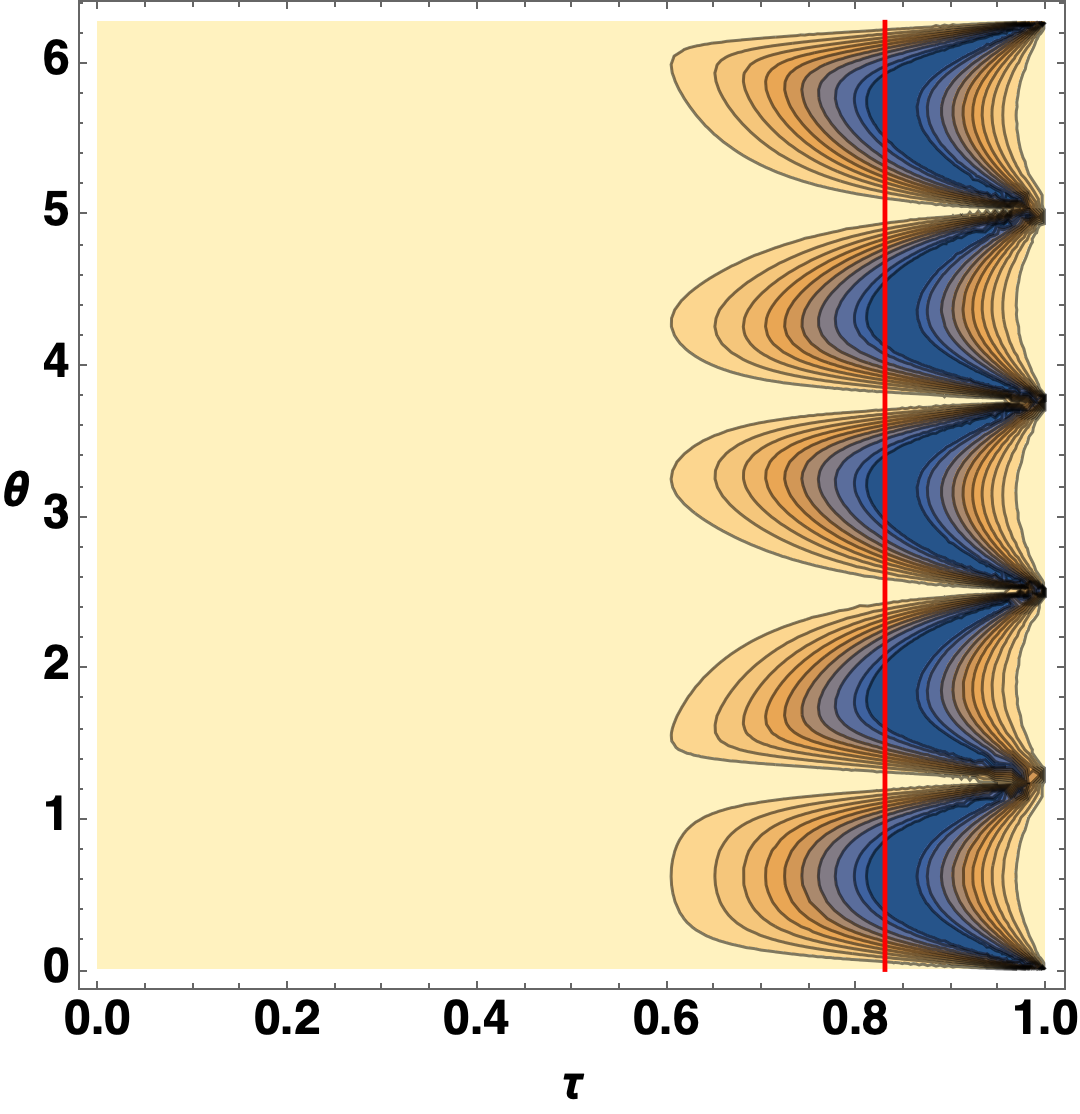}
	\caption{5 MRR Contour Plot with increments: [$0,\frac{2\pi}{5}+2\pi,\frac{4\pi}{5},\frac{6\pi}{5}+2\pi,\frac{8\pi}{5}$].}
	\label{5mrrcontour}
\end{figure}

In \Fig{5mrrcontour} we show our main 5 MRR HOMM example. Keep in mind that the choice of $\tau$ value should slice across the flattest portion of the darkest blue region at the center of the crescent-shaped contours. As you increase the number of MRRs in parallel the minimum $\tau$ value increases \cite{Kaulfuss:2023Identical}. In this case we have chosen a $\tau$ value of 0.83 as shown as the red line in \Fig{5mrrcontour}. 

For a concrete application of the HOM Multiplexer type of device, we consider an MRR array that maps the HOM effect wavelengths to the standard channels of a Dense Wavelength Division Multiplexer (DWDM). DWDMs are commonly used in classical communication and hybrid quantum-classical coexistence proposals for various quantum networking tasks \cite{Ciurana2014,Choi2014,Kleis2019}. This would allow a simple way to route NOON state outputs to distribute entanglement for quantum applications through already established classical communication networks. 

\begin{figure}[H]
    \centering
    \includegraphics[width=0.5\linewidth,keepaspectratio]{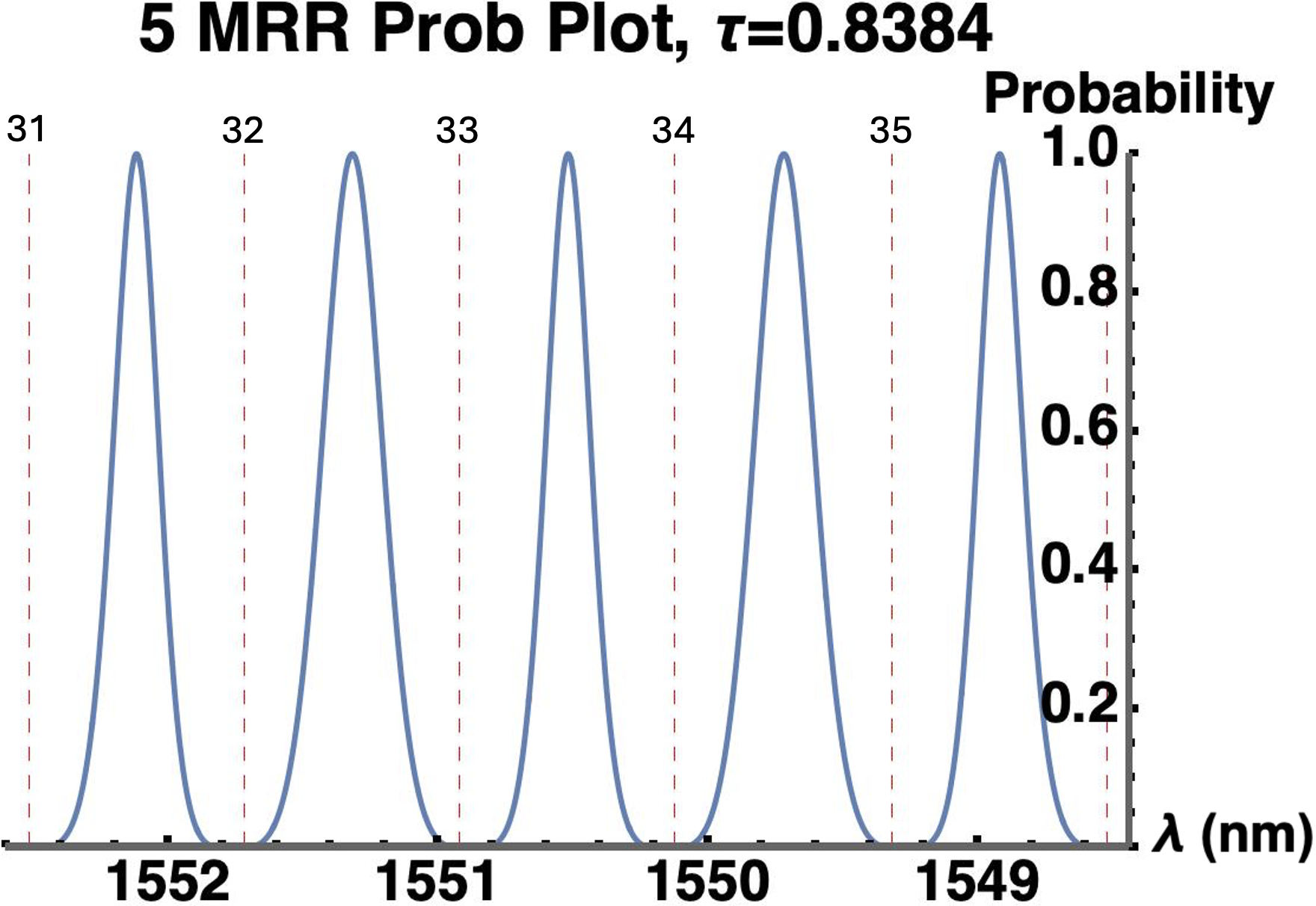}
	\caption{5 MRR Plot with increments: $[\frac{\pi}{5}, \frac{3\pi}{5}+2\pi, \frac{5\pi}{5}, \frac{7\pi}{5}+2\pi,\frac{9\pi}{5}]$ with $\tau=0.8384$. The dashed red lines are the center wavelengths for the standard 100GHz ITU DWDM channels C31-C36 and are labeled.}
	\label{DWDMfig}
\end{figure}

\Fig{DWDMfig} shows that we can manipulate the phase-shift constants on a HOM Multiplexer to place HOM effect locations (the troughs) on the standard DWDM channel central wavelengths (red dashes). Alternatively one could laterally shift the entire structure to place the peaks on the center channel wavelengths, if desired. This can be thought of as a passive, wavelength dependant device with controllable distribution of the HOM pairs. Input light with a given bandwidth (~5 nm shown in \Fig{DWDMfig}) can effectively be filtered by the Hong-Ou-Mandel Wavelength Division Multiplexer (HOMWDM) device to produce HOM pairs in specific channels.

\begin{figure}[H]
    \centering
    \includegraphics[width=0.5\linewidth,keepaspectratio]{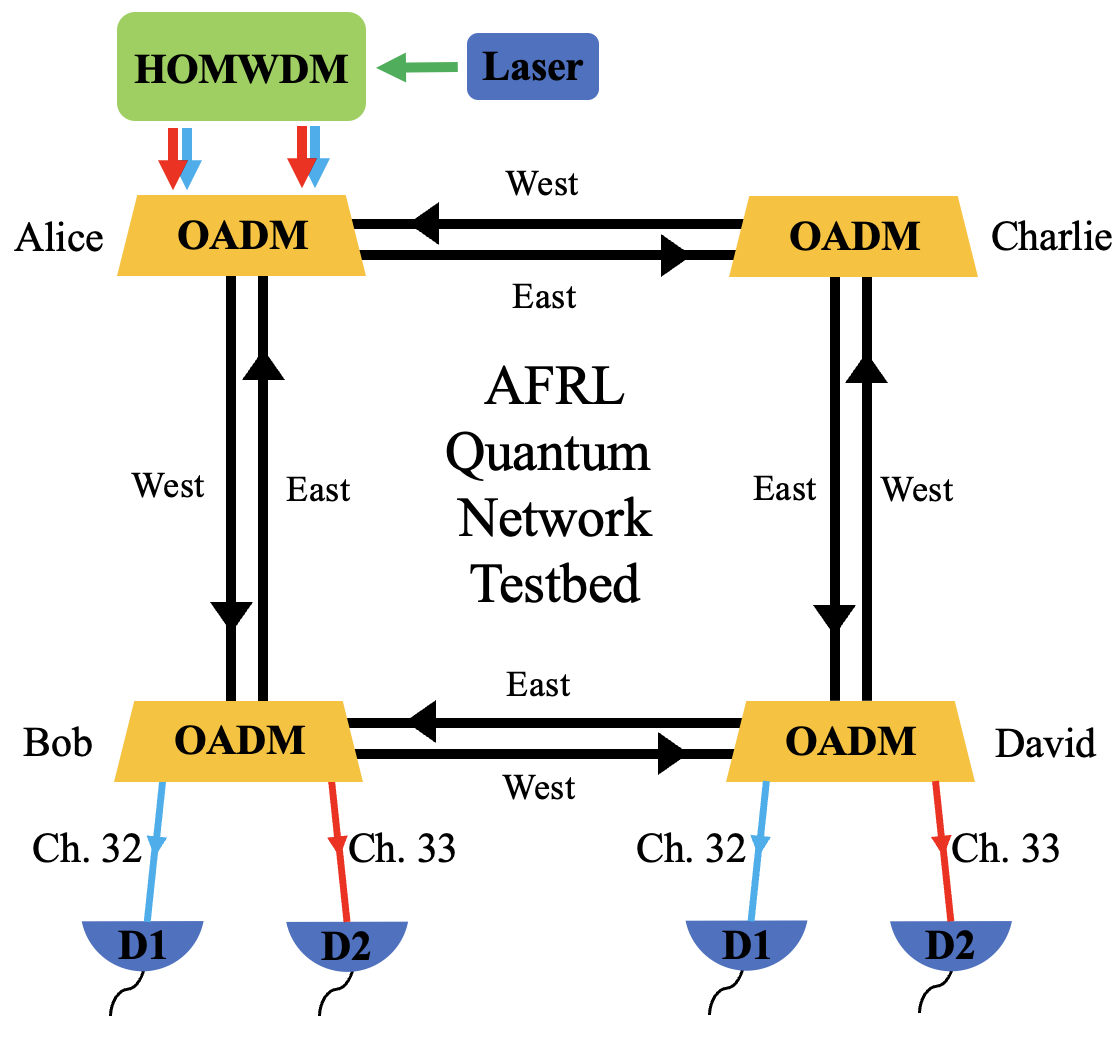}
	\caption{Schematic diagram of the AFRL Quantum Network Testbed incorporating our source for remote HOM distribution from Alice to Bob and Charlie Nodes.}
	\label{HOMWDM}
\end{figure}

To capitalize on this application we have drawn up a schematic diagram for integrating the HOM source into the AFRL Quantum Network Testbed \Fig{HOMWDM}. The network contains our source driven by a relatively broad pump at the Alice node, such that multiple DWDM channels are driven at the same time. Thus resulting in outputs that contain HOM pairs at several DWDM wavelengths as shown in Fig(\ref{DWDMfig}). Pairs are split and propagate east and west though duplex fiber in standard telecom infrastructure such as Optical Add Drop Multiplexers (OADMs). As an example this would result in HOM pairs being distributed from Alice to Bob and David (via Charlie) nodes across the network.  This represents a novel quantum resource for the AFRL Quantum Network Testbed.

\subsection*{HOM-based Entanglement Switch}
\label{switchsection}

A parallel series of regularly incremented non-identical MRRs allows for an HOM-based ``Entanglement switch'' device. This device enables the rapid swapping from a certain output of a completely separable state ($P_{1,1}=1$) to a high probability of a maximally path-entangled $\ket{2::0}$ NOON state via the HOM effect. In the `off' state the device will output completely separable states and in the `on' state the device will output a maximally entangled 2-photon state via the HOM effect. The switching of the device `off' and `on' is controlled via a shift in the round-trip phase shift of the MRRs which can be adjusted using integrated heaters. The device is scalable and easily tunable. Using the adaptability of the non-identical parallel chain of MRRs exhibited throughout this paper, the size of the shift required to switch the state of the circuit between `off' and `on' can be adjusted via the design and fabrication of the MRRs to ensure the shift is resolvable against noise for the given setup. 

\begin{figure}[H]
	\centering
	\includegraphics[width=0.5\linewidth,keepaspectratio]{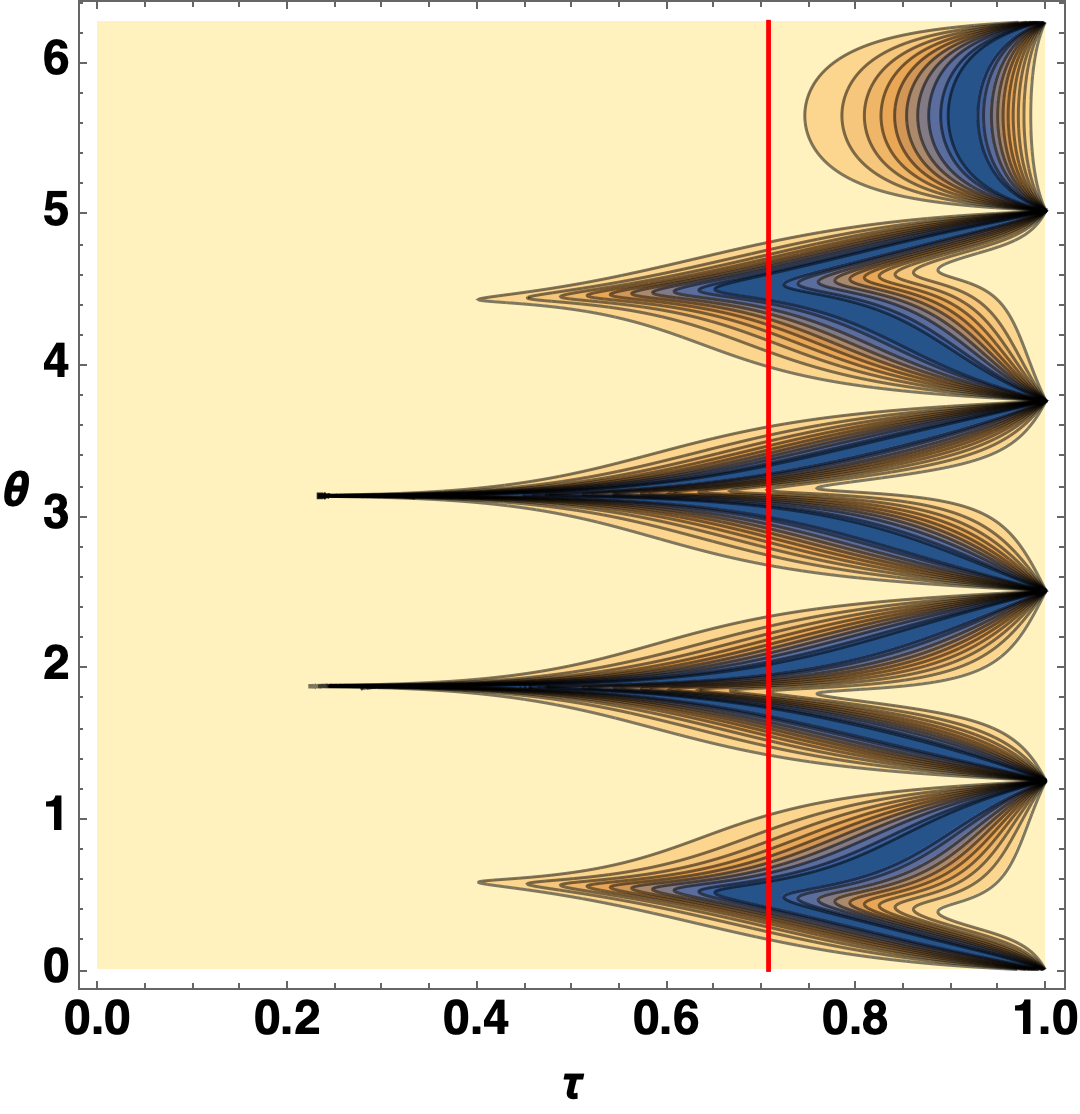}
	\caption{5 MRR HOM Switch with increments: $[-\frac{8\pi}{5},-\frac{6\pi}{5},-\frac{4\pi}{5},-\frac{2\pi}{5},0].$}
	\label{5mrrpi2.5}
\end{figure}

\begin{figure}[H]
	\centering
	\includegraphics[width=0.5\linewidth,keepaspectratio]{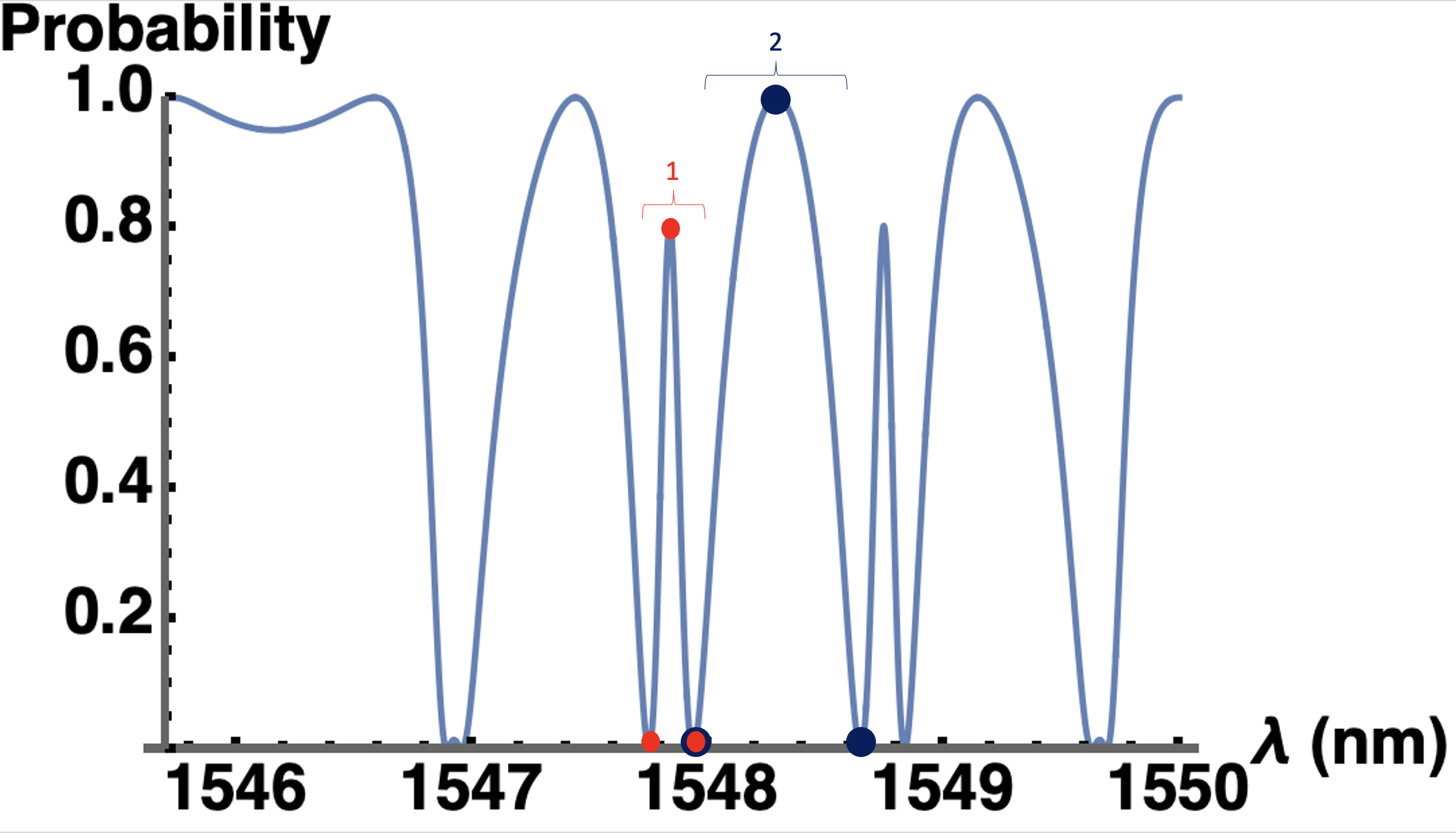}
	\caption{5 MRR HOM Switch probability plot with $\tau=\frac{1}{\sqrt{2}}$. The red points represent Switch 1 and the blue points represent Switch 2, shown in Table \ref{switchtable}. Note that both Switch 1 and Switch 2 have an overlapping point at 1547.94nm. All example switch points are given in Table \ref{switchtable}.}
	\label{5mrrpi2.5t=crit}
\end{figure}

It is important to note that in this switch case, we want the HOMM structures to be spikes, unlike the HOMWDM case, where converting to crescents is advantageous by creating a larger $\theta$ range where the HOMM condition is met. For the switch we want the $\theta$ range for the HOMM condition to be as small as resolvable against noise so that the ``switch'' between `on' and `off' states is as small in $\theta$ as possible.  

Notice between \Fig{5mrrpi2.5} and \Fig{5mrrpi2.5t=crit} that the probability plot appears reversed because increasing $\theta$ is actually decreasing $\lambda$, so the crescent structure near $\theta=2\pi$ in the contour plot appears near the low wavelengths in the probability slice, around 1546nm. 

We define the visibility ($v$) for a switch as follows:
\begin{equation}
    \label{Visibility}
    v=\frac{P_{1,1}^{max}-P_{1,1}^{min}}{P_{max}^{theory}+P_{min}^{theory}}\equiv P_{1,1}^{max}
\end{equation}

The numerator is the local maximum value of $P_{1,1}$ at the tip of the spike minus the local minimum value where $P_{1,1}=0$. The minimum value will always be zero and the denominator uses the theoretical maximum (1) and minimum (0) values; therefore, $v$ is simply the value of $P_{1,1}$ at the top of the spike for the entire switch. Therefore, the visibility value corresponds to the maximum probability of achieving completely separable states in the `off' position of the switch. $\theta$ shift is defined as the distance in $\theta$ to travel from $P_{1,1}=0$ to the maximum value where $v$ is defined. Similarly $\lambda$ shift is the corresponding shift but in terms of wavelength instead of an arbitrary $\theta$. Lastly, the $\theta$ sensitivity is the change in $\theta$ required to go from $P_{1,1}=0$ to $P_{1,1}=0.05$ and is always defined for the `on' side ($P_{1,1}=0$) of the Entanglement switch. $\theta$ shift is the amount of shift required for switch operation, and $\theta$ sensitivity informs how difficult it is to maintain the entanglement side of the switch, for $\theta$ variations larger than the $\theta$ sensitivity, $P_{1,1}$ could be greater than 0.05 and therefore not experimentally viable for the HOM effect. 

Table I describes the points we have identified as HOM switch candidates in \Fig{5mrrpi2.5t=crit}. The main identifiers of each switch are (i) the visibility ($v$), which is the maximum probability in the `on' position of the switch, and (ii) the $\theta$/$\lambda$ shift required to switch from `on' to `off'. As you can see in these examples, switch 1 requires a smaller shift in $\lambda$ to activate, but has a smaller visibility and the shift may be too small to be resolvable. This table shows that for the switches we have identified a $\theta$ shift of 0.14 rad for switch 1 and similarly 0.53 rad for switch 2 using integrated heaters would change the switch from `off' to `on'. This means for these small $\theta$ shifts the circuit output changes from completely separable single photon states to the 2 photon NOON state or vice versa. 

\begin{table}[H]
\centering
 \begin{tabular}{||c | c | c | c | c | c | c ||} 
 \hline
 Switch & $\theta$ Location (rad) & $\lambda$ Location ($nm$) & $P_{1,1}$ & Visibility ($v$) & $\theta$ Shift (rad) & $\lambda$ Shift ($nm$) \\ [0.5ex]
 \hline\hline
 \multirow{3}{1em}{\textcolor{red}{1}} & 3.31 & 1547.75 & 0 & \multirow{3}{2em}{0.80} & \multirow{3}{2em}{0.14} & \multirow{3}{2em}{0.11} \\
 \cline{2-4}
 & 3.18 & 1547.84 & 0.80 & & & \\ 
 \cline{2-4}
 & 3.04 & 1547.94 & 0 & & & \\
 \hline
 \multirow{3}{1em}{\textcolor{blue}{2}} & 3.04 & 1547.94 & 0 & \multirow{3}{2em}{1} & \multirow{3}{2em}{0.53} & \multirow{3}{2em}{0.36}\\
 \cline{2-4}
 & 2.51 & 1548.29 & 1 & & & \\
 \cline{2-4}
 & 1.99 & 1548.65 & 0 & & & \\
 \hline
\end{tabular}
\caption{Table of HOM-based Entanglement Switch values for 5 MRR non-identical series with $\frac{\pi}{2.5}$ increments, shown in \Fig{5mrrpi2.5t=crit}. Values for switch locations are given both in terms of a general $\theta$ and a specific $\lambda$ for the example values given in \Eq{lambdaeq}.}
\label{switchtable}
\end{table}

The characteristics described in this table could easily be extracted for any of the examples shown in this paper, but we have chosen an example with values we believe to be resolvable and are therefore experimentally viable representations of the HOM-based Entanglement Switch. We have verified shifts in the round trip phase as small as $\sim$3\% of the total FSR of the MRR (corresponding to a change in round-trip phase of $\Delta\theta = 0.01 \pm 0.004$ radians) \footnote{This was done using an integrated dual-MZI (Mach-Zehnder interferometer) MRR design with an FSR of $\sim$2.51nm. Telecom laser sweeps were performed using a Keysight mainframe while a Keysight N77xx detector was employed for synchronized detection. Shifts to the round-trip phase initiated  through heaters placed on the ring resonator; heating the ring changes the effective index, thereby shifting the resonant wavelengths.}. Therefore, we targeted a 5 MRR example and chose increments to yield a $\Delta\theta$ value that should be repeatable experimentally and resolvable against random fluctuations.

The HOM-based Entanglement Switch could (in theory) be customized to arbitrarily small $\theta$ ranges by continuing to add non-identical rings in series to create more spikes or by decreasing the $\theta$ offsets between the consecutive rings. For the fastest action of the switching mechanism, one should choose the smallest $\theta$ shift possible that is resolvable against noise, fluctuations, and environmental factors in the circuit experimentally. 

Overall, this should give a general understanding for the basis of the HOM-based Entanglement Switch. Further improvements and changes to design could be made to find the optimal setup depending on experimental conditions. We believe this structure could have many applications in integrated photonic circuits by allowing slight $\theta$ shifts to effectively switch the HOM effect `on' or `off'. This small shift in $\theta$ causes a drastic change in output states from a maximally path-entangled state (two photon NOON state) to a completely separable state (single photon outputs). This ability to switch rapidly between these two output modes could prove to be useful as quantum integrated photonics continue to move towards the goal of entirely `on-chip' devices from source to detection. 

\section*{Discussion}
\label{Conclusion}

In this paper we have shown the improved control over the HOMM gained by careful design and fabrication of linear chains of non-identical MRRs. This further control over the linear transformation that occurs with MRR-based devices allows for design advantages that may not be able to be replicated in tunable beamsplitters or MZI-based devices. We have shown how to manipulate the HOMM by changing the round-trip phase shift of the successive MRRs in the chain. We have also given examples of circuit designs that take advantage of the HOMM control and place the HOM effect at very specific wavelengths. Our main practical examples are the HOMWDM and the HOM-based entanglement switch. The HOMWDM shows a 5 MRR device that matches the wavelength channels of a Dense Wavelength Division Multiplexer, which allows for the distribution of HOM pairs across a network. The HOM-based entanglement switch allows for the rapid transition of the output state from $\ket{1,1}$ to $\frac{1}{\sqrt{2}}(\ket{2,0}+\ket{0,2})$. These are just two examples to show the power of chains of non-identical MRRs. The flexibility when designing and fabricating these circuits means that the non-identical MRR circuit can be precisely designed for the specific application and ready to be directly implemented ``out of the box'' without requiring any additional dynamic tuning. This could be a large advantage if, for example, one wants to operate integrated photonics at cryogenic temperatures, where integrated heaters or other external tuning methods are not feasible. Alternatively, a more general non-identical MRR circuit can be designed and dynamic tuning through on-chip heaters can be used to change the round-trip phase shifts of the MRRs in the circuit, allowing access to a wide variety of modalities from the same chip. These examples just begin to scratch the surface in the wide, higher-dimensional, solution space that can be accessed using larger chains of MRRs. Circuits of non-identical MRRs can also enhance other applications that utilize MRRs for photon routing, entanglement generation, etc., in integrated photonics. We promote these circuits of non-identical MRRs for increased control over the HOMM and encourage the further exploration of enhancements to achieving the HOM effect.   

\section*{Data availability}
Datasets are available upon reasonable requests.

\bibliography{references}

\section*{Acknowledgements}

This work was supported by the Air Force Research Laboratory through the QUEST contract (FA8750-23-C-0053).  
The US Government is authorized to reproduce and distribute reprints for Government Purposes notwithstanding any copyright notation thereon. The views and conclusions contained herein are those of the authors and do not necessarily represent the official policies or endorsements, either explicit or implied of the United States Air Force, the Air Force Research Laboratory or the U.S. Government. 

\section*{Author contributions statement}

P.K., P.A., and E.H. conceived and conducted non-identical MRR analysis. M.S. conceived the HOMWDM application. M.S., R.B., and J.S. analyzed results and provided valuable feedback. J.S. analyzed asymmetric loss and wrote section B in the supplementary information. All authors reviewed the manuscript. 

\section*{Declarations}

\subsection*{Competing Interests}
The authors declare no competing interests. 

\section*{Additional information}

Supplementary Information is available. 


\end{document}


\maketitle

\appendix
\section{Effect of $2\pi$ Increments on HOMM structures}
\label{Appendix A}

In the following equations, $\theta$ is the base MRR round trip phase shift and c1 and c2 are the constant increments added to the phase shift for MRR 1 and MRR 2, respectively. 
This two-MRR chain is the simplest possible example to see the effect of constant offsets to MRRs on the terms of the final probability function. 

\Eq{A1} is the probability for a 2 MRR chain with no offsets and \Eq{A2} is the probability for a 2 MRR chain with a $2\pi$ increment added to one of the MRRs. Notice the difference between \Eq{A1} and \Eq{A2} that occurs when a $2\pi$ increment is added to c1 (the constant addition to the phase offset of MRR 1), causing each term that contains an exponential with a half-integer multiple in the exponent to switch signs (as highlighted in blue and red). These sign changes can cause the structures of the HOMM to switch between spikes and crescents. Most terms contain integer multiples in the exponent and remain unaffected for constant additions of $2\pi$. For linear chains with more MRRs the probability terms continue to increase in complexity with the same underlying principle of half-integer multiple exponents causing HOMM structures to be converted between crescents and spikes. 

\begin{equation}
\begin{split}
P2[c1,c2,\tau,\theta]&=\left|\frac{(-1+e^{i(c1+\theta)})^2(-1+e^{i(c2+\theta)})^2\tau^4)}{(1-e^{i(c1+\theta)}\tau^2-e^{i(c2+\theta)}\tau^2+e^{i(c1+c2+2\theta)}\tau^4\highlightblue{-}e^{\frac{1}{2}i(c1+c2+2\theta)}(-1+\tau^2)^2)^2} \right. \\
&+\left. \frac{e^{i\theta}(\highlightblue{-1}+e^{\frac{1}{2}i(c1+c2+2\theta)})^2(-1+\tau^2)^2(e^{\frac{i c2}{2}}\highlightblue{+}e^{\frac{i c1}{2}}\tau^2)(\highlightblue{e^{\frac{i c1}{2}}}+e^{\frac{i c2}{2}}\tau^2)}{(1-e^{i(c1+\theta)}\tau^2-e^{i(c2+\theta)}\tau^2+e^{i(c1+c2+2\theta)}\tau^4\highlightblue{-}e^{\frac{1}{2}i(c1+c2+2\theta)}(-1+\tau^2)^2)^2}\right|^2
\label{A1}
\end{split}
\end{equation}

\begin{equation}
\begin{split}
P2[c1+2\pi,c2,\tau,\theta]&=\left|\frac{(-1+e^{i(c1+\theta)})^2(-1+e^{i(c2+\theta)})^2\tau^4)}{(1-e^{i(c1+\theta)}\tau^2-e^{i(c2+\theta)}\tau^2+e^{i(c1+c2+2\theta)}\tau^4\highlightred{+}e^{\frac{1}{2}i(c1+c2+2\theta)}(-1+\tau^2)^2)^2} \right. \\
&+ \left. \frac{e^{i\theta}(\highlightred{1}+e^{\frac{1}{2}i(c1+c2+2\theta)})^2(-1+\tau^2)^2(e^{\frac{i c2}{2}}\highlightred{-}e^{\frac{i c1}{2}}\tau^2)(\highlightred{-e^{\frac{i c1}{2}}}+e^{\frac{i c2}{2}}\tau^2)}{(1-e^{i(c1+\theta)}\tau^2-e^{i(c2+\theta)}\tau^2+e^{i(c1+c2+2\theta)}\tau^4\highlightred{+}e^{\frac{1}{2}i(c1+c2+2\theta)}(-1+\tau^2)^2)^2}\right|^2
\label{A2}
\end{split}
\end{equation}

\newpage

\section{Derivation of transmission coefficients and origin of asymmetry in HOMM interference}
\label{LossAppendix}

\subsection{Asymmetric Loss in Non-Identical Series of MRRs}

\begin{figure}[H]
	\centering
	\includegraphics[width=0.31\linewidth,keepaspectratio]{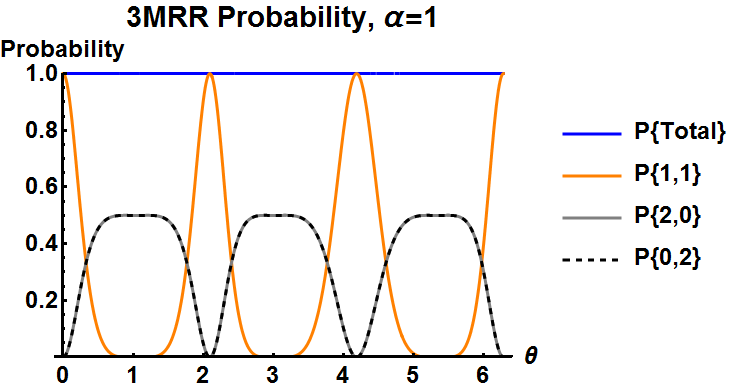}
    \includegraphics[width=0.31\linewidth,keepaspectratio]{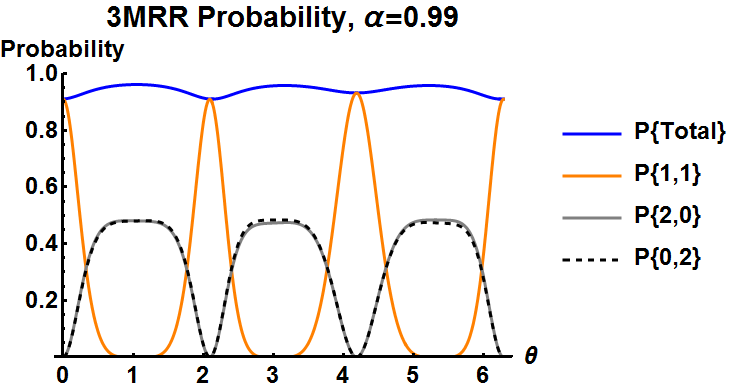}
    \includegraphics[width=0.31\linewidth,keepaspectratio]{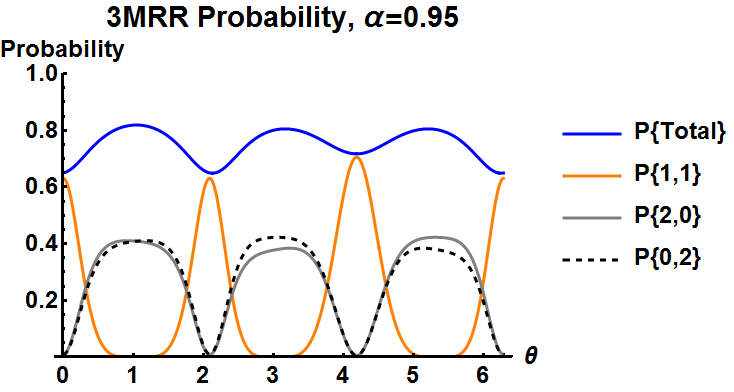}
	\caption{3 MRR full probability plots, $\alpha=1,0.99,0.95$}
	\label{3mrrloss}
\end{figure}

Full probability plots showing the probabilities of detection measured on each output state are shown for a given phenomenological round-trip radiative loss coefficient, $\alpha$, in \Fig{3mrrloss}. An $\alpha$ of 1 corresponds to exactly no loss, and the value of $1-\alpha$ gives the fractional loss, e.g. $\alpha=0.95$ corresponds to an overall $5\%$ photon loss as they travel through the rings. Notice that the loss for this non-identical system is asymmetric. This is because, at lowest order, the shortest path for photons from in/drop and add/through only touches the outermost ring of the chain. Since these MRRs are different sizes, when they have the same `$\alpha$' (a measure of loss per unit distance) the loss is also not symmetric.  

\subsection{Origin of Asymmetry}

To obtain the biphoton interference pattern in the drop and through ports of the HOMM given the halves of a photon pair are split between the input and add ports (one photon in each port), we can ultimately describe this in terms of a transfer matrix $\mathbf{T}$.
\begin{equation}
\begin{pmatrix}
\hat{a}_{in}^{\dagger} \\ \hat{e}_{in}^{\dagger}
\end{pmatrix}
\rightarrow
\begin{pmatrix}
T_{11} & T_{12} \\
T_{21} & T_{22}
\end{pmatrix}
\begin{pmatrix}
\hat{f}_{out}^{\dagger} \\ \hat{b}_{out}^{\dagger}
\end{pmatrix}
\end{equation}
Where each element of the transfer matrix is its own function of frequency. If we assume the initial state is one photon in the add port, and the other photon in the input port, we can use the transfer matrix to see how this product of modes transforms into the final output state.
\begin{align}
\hat{a}_{in}^{\dagger}\hat{e}_{in}^{\dagger}&\rightarrow (T_{11}\hat{f}_{out}^{\dagger} + T_{12}\hat{b}_{out}^{\dagger})(T_{21}\hat{f}_{out}^{\dagger} + T_{22}\hat{b}_{out}^{\dagger})\nonumber\\
&=T_{11}T_{21}(\hat{f}_{out}^{\dagger})^{2} + T_{12}T_{22}(\hat{b}_{out}^{\dagger})^{2} \nonumber\\
&\qquad +T_{11}T_{22}\hat{f}_{out}^{\dagger}\hat{b}_{out}^{\dagger} +T_{12}T_{21}\hat{b}_{out}^{\dagger}\hat{f}_{out}^{\dagger}
\end{align}
From this, we can see that the only way for the likelihood for both photons to exit the drop port to be equal to the likelihood for both photons to exit the through port, is if $T_{11}T_{21}=T_{22}T_{12}$.

\begin{figure}[H]
	\centering
	\includegraphics[width=0.95\linewidth]{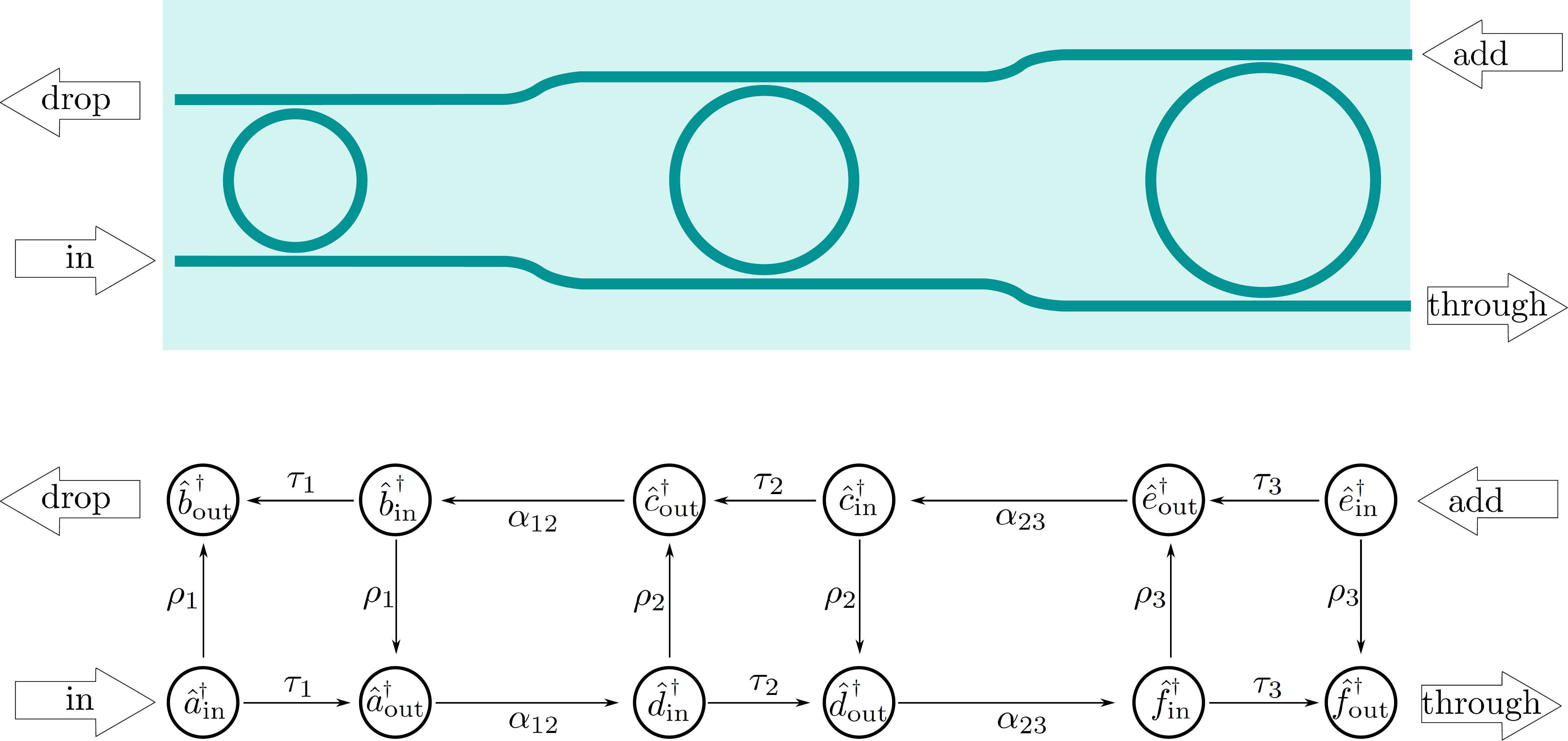}
	\caption{(Top) Diagram of three-ring configuration of HOMM system. (Bottom) Signal Flow graph representing the amplitude gains between different parts of this three-ring HOMM system.  Note: $(\tau_{1},\tau_{2},\tau_{3})$ are the overall transmission coefficients for each MRR, accounting for round-trips and all other possible paths between respective inputs and outputs, while the variable $\tau$ used earlier in this article is the coupling coefficient between a direct path from respective input to output.}
	\label{SigFlowFig}
\end{figure}

To obtain the elements of the transfer matrix, we use the signal flow graph on the bottom of Fig.~\ref{SigFlowFig}. By mirror symmetry, we already have that $T_{11}=T_{22}$ because the system when flipped vertically, maps back onto itself. Consequently, we only compute $T_{12}$ and $T_{21}$ to illustrate the asymmetry in the two-photon interference. Using the signal flow graph in Fig.~\ref{SigFlowFig}, we can use Mason's Gain Formula \cite{mason1956feedback} to compute these elements.

We first compute $T_{12}$, the overall amplitude gain from the input port $\hat{a}_{in}^{\dagger}$ to the drop port $\hat{b}_{out}^{\dagger}$.

Mason gain formula states:
\begin{equation}
T_{12}=g_{\hat{b}_{out}^{\dagger},\hat{a}_{in}^{\dagger}} = \frac{\sum_{i}G_{i}\Delta_{i}}{\Delta}
\end{equation}
Here, distinct paths have overall gains $G_{i}$ obtained by multiplying all of the in-path gains from node to node along the arrows from $\hat{a}_{in}^{\dagger}$ to $\hat{b}_{out}^{\dagger}$.

There are three principal paths from $\hat{a}_{in}^{\dagger}$ to $\hat{b}_{out}^{\dagger}$, where the light takes an overall u-turn at one of the three MRRs. The path gains are given by:
\begin{subequations}
\begin{align}
G_{1}&=\rho_{1}\\
G_{2}&=\tau_{1}^{2}\alpha_{12}^{2}\rho_{2}\\
G_{3}&=\tau_{1}^{2}\alpha_{12}^{2}\tau_{2}^{2}\alpha_{23}^{2}\rho_{3}
\end{align}
\end{subequations}

There are three principal loops within this signal flow graph, with respective loop gains given by:
\begin{subequations}
\begin{align}
L_{1}&=\rho_{1}\rho_{2}\alpha_{12}^{2}\\
L_{2}&=\rho_{2}\rho_{3}\alpha_{23}^{2}\\
L_{3}&=\rho_{1}\rho_{3}\alpha_{12}^{2}\alpha_{23}^{2}
\end{align}
\end{subequations}

The graph determinant $\Delta$ is given by:
\begin{subequations}
\begin{align}
\Delta &= 1-\sum_{i}L_{i}\;\;+\!\!\!\!\!\sum_{ij}^{\text{non-touching}}\!\!\!\!\!\!\!\!L_{i}L_{j}\;\;-\!\!\!\!\sum_{ijk}^{\text{non-touching}}\!\!\!\!\!\!\!\!L_{i}L_{j}L_{k}+...\\
&= 1-\rho_{1}\rho_{2}\alpha_{12}^{2}-\rho_{2}\rho_{3}\alpha_{23}^{2}-\rho_{1}\rho_{3}\alpha_{12}^{2}\alpha_{23}^{2}(1-\rho_{2}^{2})
\end{align}
\end{subequations}
Finally, for each path with gain $G_{i}$ there is a graph sub-determinant $\Delta_{i}$ which is equal to what $\Delta$ would be if all loops touching the path with gain $G_{i}$ were omitted in the calculation:
\begin{subequations}
\begin{align}
\Delta_{1}&=\Delta\\
\Delta_{2}&=1+\rho_{2}\rho_{3}\alpha_{23}^{2}\\
\Delta_{3}&=1
\end{align}
\end{subequations}
Altogether, this gives us for the overall value of $T_{12}$:
\begin{equation}
T_{12}= \rho_{1} +\frac{\tau_{1}^{2}\alpha_{12}^{2}\rho_{2}(1+\rho_{2}\rho_{3}\alpha_{23}^{2})}{\Delta} + \frac{\tau_{1}^{2}\alpha_{12}^{2}\tau_{2}^{2}\alpha_{23}^{2}\rho_{3}}{\Delta}
\end{equation}
Similarly, we may find the corresponding value for $T_{21}$:
\begin{equation}
T_{21}= \rho_{3} +\frac{\tau_{3}^{2}\alpha_{23}^{2}\rho_{2}(1+\rho_{2}\rho_{1}\alpha_{12}^{2})}{\Delta} + \frac{\tau_{3}^{2}\alpha_{23}^{2}\tau_{2}^{2}\alpha_{12}^{2}\rho_{1}}{\Delta}
\end{equation}
From this, we can see that in general $T_{12}\neq T_{21}$, but we can illustrate this more dramatically in the limit that $\rho_{2}$ approaches zero, making it effectively a two-ring system. In this limit:
\begin{equation}
T_{12}\!\rightarrow\rho_{1} + \frac{\tau_{1}^{2}\alpha_{12}^{2}\tau_{2}^{2}\alpha_{23}^{2}\rho_{3}}{1-\rho_{1}\rho_{3}\alpha_{12}^{2}\alpha_{23}^{2}}\;\;\text{:}\;\; T_{21}\!= \rho_{3}+ \frac{\tau_{3}^{2}\alpha_{23}^{2}\tau_{2}^{2}\alpha_{12}^{2}\rho_{1}}{1-\rho_{1}\rho_{3}\alpha_{12}^{2}\alpha_{23}^{2}}
\end{equation}
Even if we assume zero loss in the system $(\alpha_{12}=\alpha_{23}=\tau_{2}=1)$, we have:
\begin{equation}
T_{12}\rightarrow\rho_{1} + \frac{\tau_{1}^{2}\rho_{3}}{1-\rho_{1}\rho_{3}}\;\;\text{:}\;\; T_{21}= \rho_{3}+ \frac{\tau_{3}^{2}\rho_{1}}{1-\rho_{1}\rho_{3}}
\end{equation}
so that even in this lossless case $T_{12}\neq T_{21}$ unless $\rho_{1}=\rho_{3}$.

\bibliography{references}